\documentclass[floatfix,groupedaddress,twocolumn,showpacs,amsmath,pra,aps,amssymb,longbibliography,nofootinbib,superscriptaddress]{revtex4-2}

\usepackage{graphicx}
\usepackage{xcolor}
\usepackage{lipsum}
\usepackage{overpic}
\usepackage[english]{babel}
\usepackage{amsmath,amsfonts,amssymb,float}
\usepackage[titletoc,title]{appendix} 
\usepackage{soul}
\usepackage{mathtools}
\usepackage{bbold}
\usepackage[unicode]{hyperref}
\hypersetup{
	unicode=true, % nonLatin characters in Acrobat's bookmarks
	a4paper=true,
	plainpages=false,
	colorlinks=true,% false: boxed links; true: colored links
	linkcolor=blue,% color of internal links
	citecolor=blue,% color of links to bibliography
	filecolor=blue,% color of file links
	urlcolor=blue% color of external links
}
\newcommand{\rv}{{\bf r}}

\newcommand{\beq}{\begin{equation}}
\newcommand{\eeq}{\end{equation}}
\newcommand{\bea}{\begin{eqnarray}}
\newcommand{\eea}{\end{eqnarray}}
\newcommand{\BEQAL}{\begin{align}}
\newcommand{\EEQAL}{\end{align}}

\newcommand{\comment}[1]{{}}

\newcommand{\commentout}[1]{{}}

\newcommand{\ket}[1]{\ensuremath{\left\vert #1 \right\rangle}}
\begin{document}

\title{Optical excitations of Skyrmions, knotted solitons, and defects in atoms}
\author{C. D. Parmee}
\affiliation{Department of Physics, Lancaster University, Lancaster, LA1 4YB, United Kingdom}
\author{M. R. Dennis}
\affiliation{School of Physics and Astronomy, University of Birmingham, Birmingham, B15 2TT, United Kingdom}
\author{J. Ruostekoski}
\affiliation{Department of Physics, Lancaster University, Lancaster, LA1 4YB, United Kingdom}
\date{\today}

\begin{abstract}
Analogies between non-trivial topologies of matter and light have inspired numerous studies, including defect formation in structured light and topological photonic band-structures. 
Three-dimensional topological objects of localized particle-like nature attract broad interest across discipline boundaries from elementary particle physics and cosmology to condensed matter physics. 
Here we show how simple structured light beams can be transformed into optical excitations of atoms with considerably more complex topologies representing three-dimensional particle-like Skyrmions.
This construction can also be described in terms of linked Hopf maps, analogous to knotted solitons of the Skyrme-Faddeev model.
We identify the transverse polarization density current as the effective magnetic gauge potential for the Chern-Simons helicity term. 
While we prepare simpler two-dimensional baby-Skyrmions and singular defects using the traditional Stokes vectors on the Poincar\'e sphere for light, particle-like topologies can only be achieved in the full optical hypersphere description that no longer discards the variation of the total electromagnetic phase of vibration.
\end{abstract}

\maketitle

\section{Introduction}

Topologically non-trivial defects, textures, and knots have inspired physicists since the days of Kelvin~\cite{Thomson1869}. 
They are remarkably ubiquitous throughout physics, spanning a vast range of energy scales from cosmology and elementary particle physics, to superconductors, superfluidity, and liquid crystals. 
The universal nature of topological stability in such diverse areas provides unprecedented opportunities to use experimentally accessible laboratory systems as emulators even of cosmology and high-energy physics where the experimental evidence is absent~\cite{volovik}. 
In recent years, the experimental study of topological defects and textures in structured optical fields has emerged as one of the most promising areas to engineer and detect topologically non-trivial characteristics~\cite{strlight_review}, including singularities of the phase or polarization that may form knotted or linked  geometries~\cite{Leach2004,Dennis2010,Kedia2013,Larocque2018} or M\"obius strips~\cite{Bauer964}. Another line of research on non-trivial topologies of light has focused on photonic band structures~\cite{Ozawa_review}, analogous to electronic band structures in crystals.

Topological Skyrmionic textures~\cite{Skyrme1961} are non-singular, localized spin (or pseudo-spin) configurations that do not perturb the spin profile sufficiently far from the center of the structure. 
The non-trivial topology of the object arises from how the spatial profile of the spin texture wraps over the spin configuration, or order-parameter, space. 
Particle-like Skyrmions, defined by mappings of the $\Pi_3$ homotopy group, display non-trivial three-dimensional (3D) spatial profiles and are well-known in nuclear and elementary particle physics as theoretical paradigms~\cite{manton-sutcliffe, Battye97,Battye09,donoghue2014}. 
Similar structures have been proposed in cosmological models~\cite{radu_physrep_2008}, and they have also been actively investigated in superfluids~\cite{Volovik1977,Shankar1977, Ruostekoski2001, al-khawaja_nature_2001}, with a particular focus on their energetic stability~\cite{battye_prl_2002,savage_prl_2003, Ruostekoski2004, Kawakami12,Tiurev_2018}. 
Hopfions, classified by the integer-valued Hopf charge, are closely related to the 3D Skyrmions and have attracted particular attention owing to their tendency to form stable torus knots~\cite{faddeev_nature_1997, Battye98, Sutcliffe17, HIETARINTA99, Babaev2002}. 
Although 3D Skyrmions and Hopfions have recently been experimentally realized as stationary superfluid configurations~\cite{Hall2016, Lee2018} in liquid crystals~\cite{Ackerman2017}, and in phase and polarization-structured light~\cite{Sugic2021}, the wider attention has focused on much simpler, planar analogs, 2D baby-Skyrmions. 
2D baby-Skyrmions, like their 1D cousins~\cite{borgh_prl_2016a}, exhibit topologically non-trivial, non-singular configurations in reduced dimensions and have the best known early examples in superfluids as the Anderson-Toulouse-Chechetkin~\cite{anderson_prl_1977, chechetkin_jetp_1976} and Mermin-Ho~\cite{mermin_prl_1976} non-singular vortices, owing to their ability to carry angular momentum. 
A large body of more recent research has included magnetic systems~\cite{muhlbauer_science_2009,Nagaosa2013}, with potential data storage applications, rotating atomic superfluids~\cite{leanhardt_prl_2003, Leslie2009, choi_prl_2012,weiss_ncomm_2019,ho_prl_1998,mizushima_prl_2002,Lovegrove2014}, exciton-polariton structures~\cite{Cilibrizzi2016, Donati2016, Krol2021}, and optical fields~\cite{Tsesses2018, Du2019, Gao2020, Davis2020,Gutierrez-Cuevas2021}.

For the particular case of baby-Skyrmions in optical fields, field profiles are usually analyzed using the Stokes vector, i.e., a point on the Poincar\'e sphere, corresponding to the coherent, transverse polarization state at each point in the field. 
However, going beyond these more easily observable parameters, the full topology of the field configurations, crucially, also depends on the spatial variation of the total phase of vibration on the polarization ellipse~\cite{Bliokh2019}, which is the sum of the phases of the electric field components, and is not represented by the Stokes vector. 
This complete topology is then described by the optical hypersphere $S^3$ (unit sphere in 4D)~\cite{Sugic2021}, allowing, e.g., for full 3D particle-like topologies of light.

Here we utilize simple configurations of structured light fields to show how these can lead to optical excitations in atomic media of comparable or considerably more complex topologies. 
Baby-Skyrmions, represented by full Poincar\'e beams~\cite{Beckley2010} in light fields, can straightforwardly be transferred to optical excitations, and therefore frozen and stored, in strongly confined oblate atomic ensembles. We consider a $J=0\rightarrow J'=1$ transition that can form, e.g., in $^{88}$Sr very long-lived excitations. 
By going beyond the Stokes representation of light beams to incorporate the full degrees of freedom of the field amplitudes, where we no longer discard the spatial variation of the sum of the phases for the two field components, we can form 3D particle-like Skyrmions, localized in space. 
We identify the transverse polarization density of the atoms as a synthetic magnetic vector potential of the 3D Skyrmions with non-trivial helicity.
While constructing such an object directly in a light beam is quite challenging even for modern structured light engineering~\cite{Sugic2021}, we show how appropriately adjusting the light-matter coupling provides a solution with simple copropagating beams.  
For this solution, we then formulate the Stokes representation to provide precisely a Hopf fibration between the optical hypersphere and the Poincar\'e sphere, representing knotted solitons or Hopfions, analogous to the knotted solitons in the Skyrme-Faddeev model~\cite{faddeev_nature_1997,manton-sutcliffe}, and show the linked and trefoil knot Hopfion preimages of the Poincar\'e sphere.
While such objects are non-singular, we also show how singular defects can be transferred from light to optical excitations. For systems where the light scattering is strong, and light mediates dipole-dipole interactions between the atoms, we remarkably find that singular defects can even exist as collective excitation eigenmodes.
These behave as spatially delocalized `superatoms', exhibiting their own collective resonance linewidth and line shift.

\section{Baby-Skyrmions} \label{baby}

We first show how to prepare 2D baby-Skyrmions in an atomic ensemble.
A non-singular topological texture can be constructed by letting a (pseudo-)spin orient into a localized structure that points in every direction somewhere within a 2D plane, but takes a uniform constant value everywhere sufficiently far away from the origin, independently of the direction. 
The plane can then be compactified to a unit sphere $S^2$ and the orientations of the spin on the 2D plane can be characterized by $S^2\rightarrow S^2$ mappings. 
Such mappings can take topologically non-trivial values, associated with the existence of baby-Skyrmions, also frequently called non-singular vortices. 

For optical fields, the state is most commonly characterized on the $S^2$ Poincar\'e sphere by an easily observable Stokes vector $\textbf{S}$~\cite{BOR99,Bliokh2019}, and the $S^2\rightarrow S^2$ mapping defining the baby-Skyrmion topology counts the number of times the object wraps over $S^2$,
\begin{equation}\label{Eq:SkyrmionNumber}
W =\int_{\mathcal{S}}\frac{d\Omega_i}{8\pi} \epsilon_{ijk} \textbf{S}\cdot\frac{\partial\textbf{S}}{\partial r_j}\times \frac{\partial \textbf{S}}{\partial r_k},
\end{equation}
where $\epsilon_{ijk}$ denotes a completely antisymmetric Levi-Civita tensor.
A field configuration that satisfies a non-trivial winding $W = 1$ can be achieved using a superposition of a Gaussian and Laguerre-Gaussian (LG) beam, with wavevector $\textbf{k}$ and frequency $\omega = c|\textbf{k}|=ck$. 
Working with slowly-varying amplitudes for the light and atoms by factoring out the fast-rotating term $\exp(-\text{i}\omega t)$, the positive frequency component of the field, $\boldsymbol{\mathcal{E}}{}(\textbf{r})$, is given by
\begin{equation}\label{Eq:BabySkyrmionField}
\boldsymbol{\mathcal{E}}{}(\textbf{r})=\text{U}_{0,0}(w_0)\hat{\textbf{e}}_x+\text{U}_{1,0}(w_0)\hat{\textbf{e}}_y.
\end{equation}
Here $\text{U}_{l,p}(w_0)$ are the LG modes with azimuthal quantum number $l$, radial quantum number $p$, and focused beam width $w_0$~\cite{strlight_review}.
The light field of Eq.~\eqref{Eq:BabySkyrmionField}, now a full Poincar\'e beam~\cite{Beckley2010}, contains a N\'eel type baby-Skyrmion whose optical polarization we have defined here in the linear $\hat{\textbf{e}}_{x,y}$ basis, instead of the commonly used circular basis~\cite{Donati2016,Gao2020,Gutierrez-Cuevas2021}, because the linear basis is physically relevant when manipulating the atomic transition, as discussed in Sec.~\ref{Skyrmions}.

Topologically non-trivial fields in this simple example can be straightforwardly transformed to optical excitations in atomic ensembles. We consider a $\ket{J=0, m=0}\rightarrow\ket{J' = 1, m=\upsilon}$ transition which can in alkaline-earth-metal-like atoms be very narrow, forming long-lived excitations. For instance, the $^{88}$Sr clock transition ${}^1S_0\rightarrow {}^3P_0$ has a linewidth controllable by a magnetic field, with the transition entirely forbidden at the zero field.
We create a non-singular topological texture of the optical excitation by considering an oblate ensemble of atoms, strongly confined along the light propagation direction ($z$ axis).
We write the optical excitation as an electric polarization density, or the density of electromagnetic vibration in atoms, with the slowly-varying positive frequency component $\textbf{P}{}(\textbf{r})= \sum_j\delta(\textbf{r}-\textbf{r}_j)\textbf{d}_j$. The induced dipole $\textbf{d}_j = \mathcal{D}\sum_{\upsilon}\hat{\textbf{e}}_{\upsilon}\mathcal{P}^{(j)}_{\upsilon}$ on atom $j$, located at $\textbf{r}_j$, is given in terms of the reduced dipole matrix element $\mathcal{D}$ and the excitation amplitudes $\mathcal{P}_{\upsilon}^{(j)}$, with the unit vectors $\hat{\textbf{e}}_{\pm}=\mp(\hat{\textbf{e}}_{x}\pm\text{i}\hat{\textbf{e}}_{y})/\sqrt{2}$ and $\hat{\textbf{e}}_{0}=\hat{\textbf{e}}_{z}$.
Light couples to $\textbf{P}$ via the atomic polarizability, $\alpha = -\mathcal{D}^2/[\hbar\epsilon_0(\Delta +\text{i}\gamma)]$, according to $\textbf{P}{}(\textbf{r})= \epsilon_0\alpha\boldsymbol{\mathcal{E}}{}(\textbf{r})$, where $\gamma$ denotes the resonance linewidth of the atom and $\Delta$ is the detuning of the laser frequency from the atomic resonance. The excitation is then re-emitted back to the light field, where the scattered light amplitude is given by 
$\int d^3 r' \mathsf{G}(\textbf{r}-\textbf{r}') \textbf{P}{}(\textbf{r}')$ and $\mathsf{G}(\rv)\textbf{d}$ denotes the dipole radiation at $\rv$ from an oscillating dipole $\textbf{d}$ at the origin~\cite{Jackson}.

\begin{figure}
	\hspace*{0cm}
	\includegraphics[width=0.9\columnwidth]{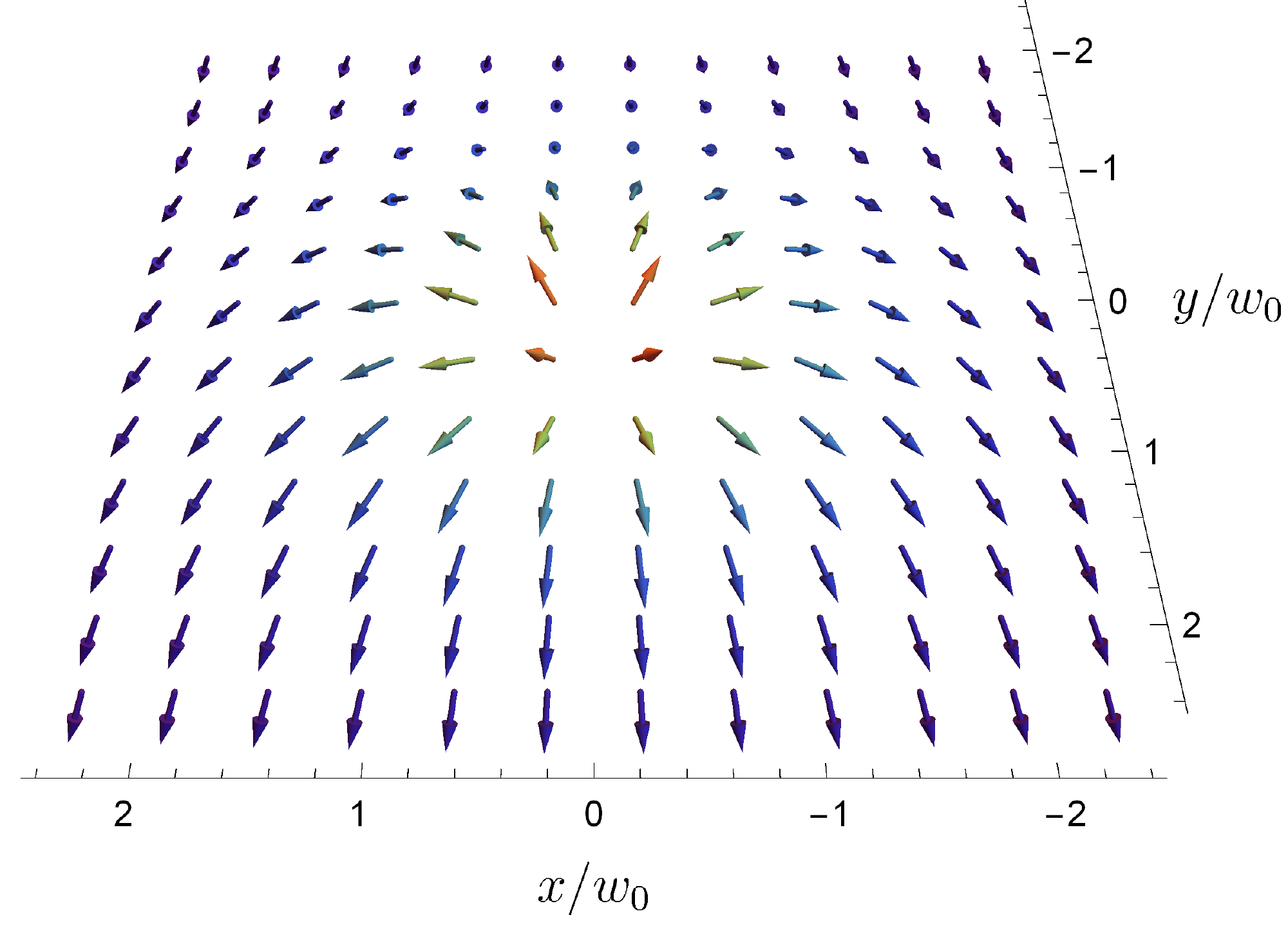}
	\vspace{-0.2cm}
	\caption{Optical excitation of a baby-Skyrmion with a characteristic fountain-like structure in the atomic Stokes vector $\textbf{S}$ [Eq.~\eqref{Eq:StokesParamAtoms}], generated by the light field of Eq.~\eqref{Eq:BabySkyrmionField}. For illustrative purposes, we choose a regular square array, but any geometry can be chosen.
	The vector coloring corresponds to the $S_3$ component of $\textbf{S}$.}
	\label{Fig:Model}
\end{figure}

For describing the topology of the optical excitation, we define a pseudo-spinor in terms of the normalized transverse atomic polarization densities,
\beq
\hat{\sf P}(\rv)= 
\begin{pmatrix}
\hat P_x (\rv) \\
\hat P_y (\rv)
\end{pmatrix},
\eeq
where, as for the light field in Eq.~\eqref{Eq:BabySkyrmionField}, we work in a linear rather than circular basis, with $\hat P_j=P_j/|\textbf{P}|$ and the longitudinal component $P_z=0$.
We can then define the corresponding atomic Stokes vector 
\begin{equation}\label{Eq:StokesParamAtoms}
S_j(\textbf{r})=\hat{\mathsf{P}}^{\dagger}\sigma_{j}\hat{\mathsf{P}},
\end{equation}
where $\sigma_j$ are the Pauli matrices.
In Fig.~\ref{Fig:Model}, we show the baby-Skyrmion configuration generated by the field in Eq.~\eqref{Eq:BabySkyrmionField}. The atomic Stokes vector, Eq.~\eqref{Eq:StokesParamAtoms}, now has a fountain-like structure, $\textbf{S}=[2\sqrt{2}\rho w_0\hat{\textbf{e}}_{\rho}+(w_0^2-2\rho^{2})\hat{\textbf{e}}_z]/(w_0^2+2\rho^{2})$, and takes a uniform value $\textbf{S}=(0,0,-1)$ sufficiently far away from the center of the object.
It is easy to verify that the winding number Eq.~\eqref{Eq:SkyrmionNumber} for $\textbf{S}$ integrates to $W=1$, and that the same topological structure in the incident field is excited in the atomic polarization density. The principle of creating a baby-Skyrmion is therefore closely related to the studies of analogous objects in exciton-polariton systems~\cite{Cilibrizzi2016, Donati2016}.

\section{Particle-like objects}\label{Sec:Solitons}

\subsection{3D Skyrmions}\label{Skyrmions}

We now show how 3D Skyrmionic structures can be constructed by considering the full complex nature of the electric polarization.
The Stokes vector representation of the Poincar\'e sphere for the light field amplitudes or optical excitations in atoms [Eq.~\eqref{Eq:StokesParamAtoms}] does not provide the full field description, as the texture may also exhibit non-trivial, non-uniform spatial variation of the total phase of the two field components, which is discarded. 
A more complete description of the field topology can instead be obtained using the optical hypersphere $S^3$ \cite{Sugic2021}. 

The field parametrization in $S^3$ permits considerably more complex, particle-like objects, localized in 3D physical space. 
Compactifying the real 3D space, such that the fields are assumed to take the same value far away from the particle, independently of the direction, allows us to describe the topology by $S^3\rightarrow S^3$ mappings. 
Such mappings can be characterized by distinct topological equivalence classes, identified by the third homotopy group elements $\Pi_3(S^3)=\mathbb{Z}$.
Non-trivial objects whose $S^3$ mappings wrap over the order parameter space an integer number of times represent topologically non-trivial solutions, originally introduced by Skyrme~\cite{Skyrme1961}.

We now parametrize the atomic polarization spinor on the $S^3$ optical hypersphere by writing it as a four-component unit vector $\hat{\textbf{n}} = (n_1,n_2,n_3,n_4)$, and taking
\begin{equation}\label{Eq:ndefn}
\hat{\mathsf{P}} = \begin{pmatrix} n_2 + \text{i}n_1 \\ n_4 + \text{i}n_3 \end{pmatrix}= \begin{pmatrix}\text{i}\sin\psi \sin\beta\exp(-\text{i}\eta) \\ \cos\psi+\text{i}\sin\psi \cos\beta  \end{pmatrix},
\end{equation}
where $\hat{\textbf{n}}$ is represented by the hyperspherical angles $0<\psi,\beta \leq \pi$ and $0<\eta\leq2\pi$.
The integer topological charge of the 3D Skyrmion (known in high-energy physics as the baryon number~\cite{donoghue2014}), is found then by counting the number of times $\hat{\textbf{n}}$ wraps over $S^3$,
\begin{equation}\label{Eq:3DSkyrmionNumber}
B =\int d^3r \mathcal{B}(\textbf{r})  =-\int \frac{d^3r}{2\pi^2}\epsilon_{ijk}\epsilon_{a b c d} n_a\frac{\partial n_b}{\partial r_i}\frac{\partial n_c}{\partial r_j}\frac{\partial n_d}{\partial r_k},
\end{equation}
where $\mathcal{B}(\textbf{r})$ is the topological charge density.
By introducing the transverse polarization density current
$\textbf{J}= \frac{1}{2{\text{i}}}[{ \hat{\sf P}}^\dagger \nabla {\hat{\sf P}}- (\nabla {\hat{\sf P}}^\dagger){\hat{\sf P}}]$, $\mathcal{B}$ can be rewritten as 
\begin{equation}\label{Eq:OpticalCurrent}
\mathcal{B}(\textbf{r}) = -\frac{1}{4\pi^2}\textbf{J}\cdot\nabla\times\textbf{J},
\end{equation} 
and is therefore analogous to the linking number density in (super)fluids~\cite{Volovik1977}, where $\textbf{J}$ is replaced by the (super)fluid velocity, and to the Chern-Simons term for the magnetic helicity~\cite{Jackiw00}, in which case $\textbf{J}$ represents the gauge potential for the magnetic field (Note that the sign of the winding numbers may vary depending on the orientations of the coordinates and the mappings).
To understand the structure of the Skyrmion in Eq.~\eqref{Eq:ndefn}, we consider a simple analytic mapping from 3D Euclidean real space to the optical hypersphere~\cite{Ruostekoski2004} with $\eta = p\phi$, $\beta = \theta$ and $\psi = q\varsigma(r)$, finding that Eq.~\eqref{Eq:3DSkyrmionNumber} integrates to give a topological charge $B=pq$, where the monotonic function $\varsigma(r)$ satisfies $\varsigma(0)=0$ and $\varsigma\rightarrow \pi$ sufficiently far from the origin.
The first spinor component vanishes along the $z$ axis, and now forms a multiply-quantized vortex line with a winding number $p$. 
The second component vanishes at the circles $\theta = \pi/2$, $r = \varsigma^{-1}[(n-1/2)\pi/q]$ for $n=1,\ldots,q$, with $\hat{P}_y \sim -\delta r - \text{i}\delta\theta $ in the circle vicinity, and hence forms $q$ concentric vortex rings with different radii.
The vortex line threads the vortex rings, and has a non-vanishing density confined inside the toroidal regions around the vortex ring singularities, such that the Skyrmion is spatially localized, forming a particle-like object.
Any continuous deformation of Eq.~\eqref{Eq:ndefn} conserves the discrete topological charge; a 3D Skyrmion with $B=pq$ can also be constructed by taking any combination of singly- and multiply-quantized lines (rings) with total winding $q$ ($p$), located in the components of $P_x$ ($P_y$), where the lines thread through the rings.

Forming such a structure in the polarization density using electromagnetic fields in free space alone is a rather challenging task of structured light engineering~\cite{Sugic2021}.
However, we can here exploit the properties of the light-matter coupling to simplify the field profiles considerably. 
To create the Skyrmion, we take a coherent superposition of copropagating light beams
\begin{equation}\label{Eq:SkyrmionField}
 \boldsymbol{\mathcal{E}}{}(\textbf{r})=\text{U}_{l,0}(w_x)\hat{\textbf{e}}_x+[\text{U}_{0,0}(w_1)-c\text{U}_{0,0}(w_2)]\hat{\textbf{e}}_y,
\end{equation}
where for the LG beam we now choose $l=1$ to form a $B=1$ Skyrmion, although we consider higher-order charges in the next section. For the Gaussian beams of unequal focusing, the parameter $c=\exp(-\rho_0^2/w_1^2+\rho_0^2/w_2^2)$ defines the circular radius $\rho_0$ in the $z=0$ plane of minimum focusing at which they interfere destructively. Destructive interference outside the ring is prevented due to diffraction. Diffraction also leads to variation of the phase (Gouy phase), such that $\text{U}_{0,0}(w_1)-c\text{U}_{0,0}(w_2) \sim (\rho-\rho_0)+\text{i}\zeta z$ in the zero field ring vicinity. The $y$-polarized light component now forms a singular vortex ring~\cite{Ruostekoski2005} with a $2\pi$
phase winding, analogously to $\hat P_y$ of Eq.~\eqref{Eq:ndefn} for $q=-1$, and a vortex core anisotropy
\begin{equation}\label{Eq:Anisotropy}
\zeta = \frac{w_1^2w_2^2-\rho_0^2(w_1^2+w_2^2)}{w_1^2w_2^2\rho_0k}.
\end{equation}
The $x$-polarized light component exhibits a singular vortex line, analogously to $\hat P_x$ of Eq.~\eqref{Eq:ndefn} for $p=-1$, where the LG beam has an intensity that reaches its maximum in the $z=0$ plane at $\rho = w_x/\sqrt{2}$, coinciding with the vortex ring singularity. However, the intensity is not confined along the $z$ direction as required by the Skyrmion solution Eq.~\eqref{Eq:ndefn}. 
In order to achieve the desired profile, we can utilize the light-matter coupling, which can be selectively turned on around the $z=0$ plane only, to confine $P_x$. This can be achieved by controlling the $m=0$ quadratic Zeeman level shift, either by magnetic fields, or ac Stark shifts of lasers or microwaves~\cite{gerbier_pra_2006}.

\begin{figure*}
	\hspace*{0cm}
	\includegraphics[width=2\columnwidth]{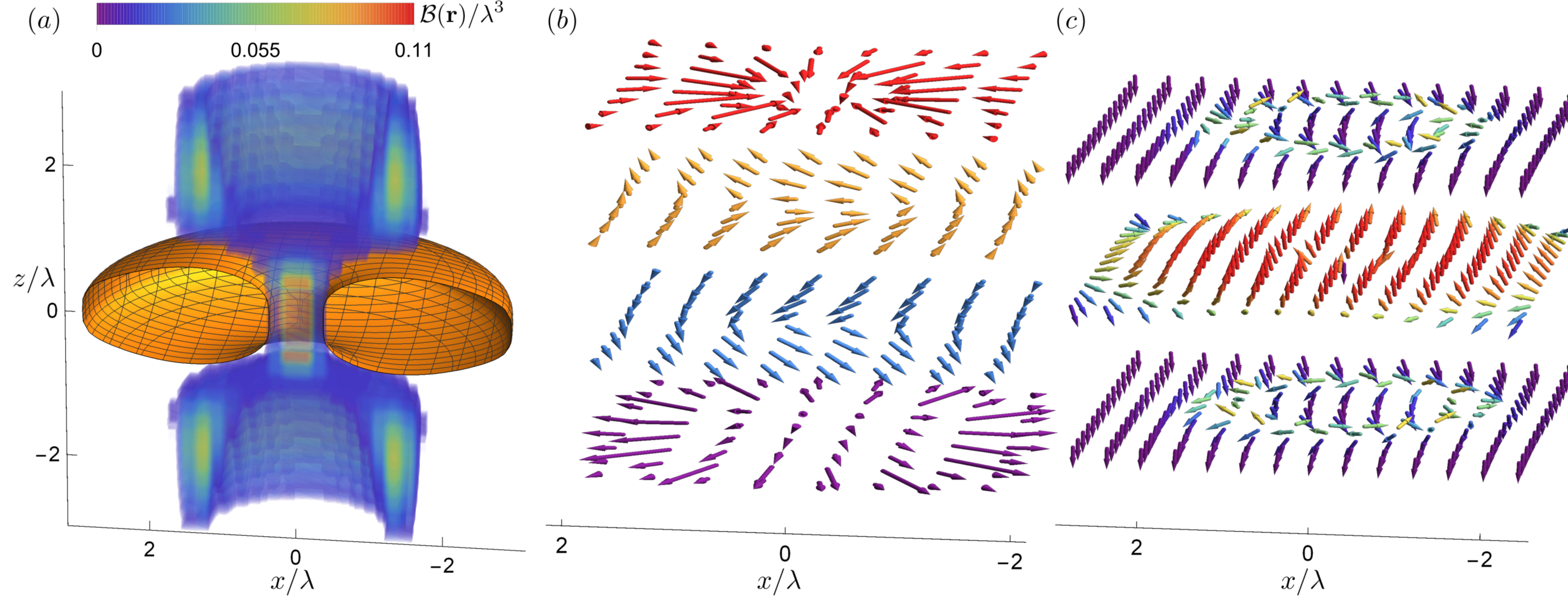}
	\vspace{-0.2cm}
	\caption{Topological particle-like objects of optical excitations: (a) a 3D Skyrmion with the topological charge $B=1$, prepared using the light field in Eq.~\eqref{Eq:SkyrmionField} for $l=1$; (b) its artificial gauge vector potential field, represented by the polarization density current $\textbf{J}$; and (c) the construction of the Hopfion field profile $\hat{\textbf{h}}$.
	(a) The isosurface of $|P_x|^2 = 0.6 \mathcal{D}^4 |\boldsymbol{\mathcal{E}}{}(\textbf{0})|^2/(\hbar \gamma)^2$ (meshed region) shows the confinement of the $x$-polarized electromagnetic vibration of the atoms due to the applied level shifts.
	The topological charge density $\mathcal{B}(\textbf{r})$ (colored region) is concentrated at the origin and in two rings where the gradients of the atomic polarization density become large, with the corresponding $\textbf{J}$ in (b) exhibiting a flow in the $\pm x$ directions near the origin or radial flow at the charge density rings.
	In (b, c), a geometry of stacked square arrays is chosen for illustrative purposes, although any geometry can be used. The vector coloring in (c) corresponds to the $h_3$ component of $\hat{\textbf{h}}$.
	The beam widths $(w_x,w_1,w_2)/\lambda = (2,3,4.5)$.
	}
	\label{Fig:ChargeDensity}
\end{figure*}

In Fig.~\ref{Fig:ChargeDensity}(a), we show the topological charge density $\mathcal{B}$ for the 3D Skyrmion constructed using the field in Eq.~\eqref{Eq:SkyrmionField}, [where we have ignored any contribution from the beam phase factor $\exp(\text{i}kz)$], and the confinement of $P_x$, achieved using spatially dependent level shifts $\Delta_x(\textbf{r}) =\delta[1-\exp(-z^2/10w_x^2)]$ in $\textbf{P}{}(\textbf{r})=\epsilon_0\alpha(\Delta_{\upsilon}) \boldsymbol{\mathcal{E}}{}(\textbf{r})$. We consider long-lived excitations with extremely narrow linewidth, so typically $\delta\gg\gamma$, and we take
$\delta/\gamma=200$.
The topological charge density shows the localization of the Skyrmion, with the density concentrated at the origin, and also in two rings where the gradient of $P_x$ and $P_y$ becomes large from the applied level shifts and vortex ring phase winding, respectively.
Changing the vortex ring core anisotropy, Eq.~\eqref{Eq:Anisotropy}, which has the value $\zeta = 0.08$ in Fig.~\ref{Fig:ChargeDensity}(a), increases the concentration in the rings for a more anisotropic core.
We find the corresponding transverse polarization density current $\textbf{J}$ [Fig.~\ref{Fig:ChargeDensity}(b)], which represents the synthetic magnetic vector potential with an integer linking number, has a large magnitude where the charge density is highly concentrated. At the charge density rings, $\textbf{J}$ flows radially inwards or outwards, while closer to the origin, $\textbf{J}$ flows almost entirely along the $\pm x$ directions.

\subsection{Knotted solitons}\label{KnottedSolitons}

We have shown how 3D particle-like objects can be prepared by going beyond the Stokes vector representation used to describe baby-Skyrmions in Sec.~\ref{baby} and parametrizing the optical excitations on $S^3$. 
However, we can also construct particle-like 3D objects using the Poincar\'e sphere, instead of the full optical hypersphere. 
The advantages of our choice of representation for the optical hypersphere in Eq.~\eqref{Eq:ndefn} become apparent when we formulate the $S^3\rightarrow S^2$ transformation from the optical hypersphere to the Stokes vector precisely as a Hopf fibration~\cite{Hopf1931,Urbantke2003,Sugic2021}. 

The Hopf fibration, initially of purely mathematical interest, arises naturally in field theories. 
In the Skyrme-Faddeev model, 3D topological objects known as Hopfions are classified by an integer-valued Hopf charge~\cite{faddeev_nature_1997,Battye98,Sutcliffe17,HIETARINTA99,Babaev2002}. Considerable interest in these systems was generated by the observations that the stable solutions may exhibit knots. 
The Hopf map of the vector $\hat{\textbf{n}}$ on $S^3$ to a vector $\hat{\textbf{h}}=(h_1,h_2,h_3)$ on $S^2$  is given by 
\begin{subequations}\label{Eq:HopfionMapping}
\begin{align}
&h_1 = 2(n_1n_3+n_2n_4),\\
&h_2 = 2(n_2n_3-n_1n_4),\\
&h_3 = n_1^2+n_2^2-n_3^2-n_4^2,
\end{align}
\end{subequations}
where the mapping falls into distinct topological equivalence classes $\Pi_3(S^2)=\mathbb{Z}$, characterized by the integer Hopf charge, $Q_H$.
Upon substituting the expressions for $\hat{\textbf{n}}$ in terms of $\hat{\mathsf{P}}$, the mapping indeed returns the atomic Stokes vector, Eq.~\eqref{Eq:StokesParamAtoms}.
Applying the Hopf map of Eq.~\eqref{Eq:HopfionMapping} to the $B=1$ Skyrmion of Fig.~\ref{Fig:ChargeDensity}(a), we obtain a Hopfion with charge $Q_H=1$, shown in Fig.~\ref{Fig:ChargeDensity}(c) by the field profile $\hat{\textbf{h}}$.
The particle-like nature of the Hopfion is clearly visible, where the full 3D spin texture is localized around the origin. 
At large distances in any direction from the center, and along the vortex line where $\hat{P}_x$ vanishes, we have $\hat{\textbf{h}}=(0,0,-1)$, while at the vortex ring with $\hat{P}_y=0$, $\hat{\textbf{h}}=(0,0,1)$.

\begin{figure}
	\hspace*{0cm}
	\includegraphics[width=0.9\columnwidth]{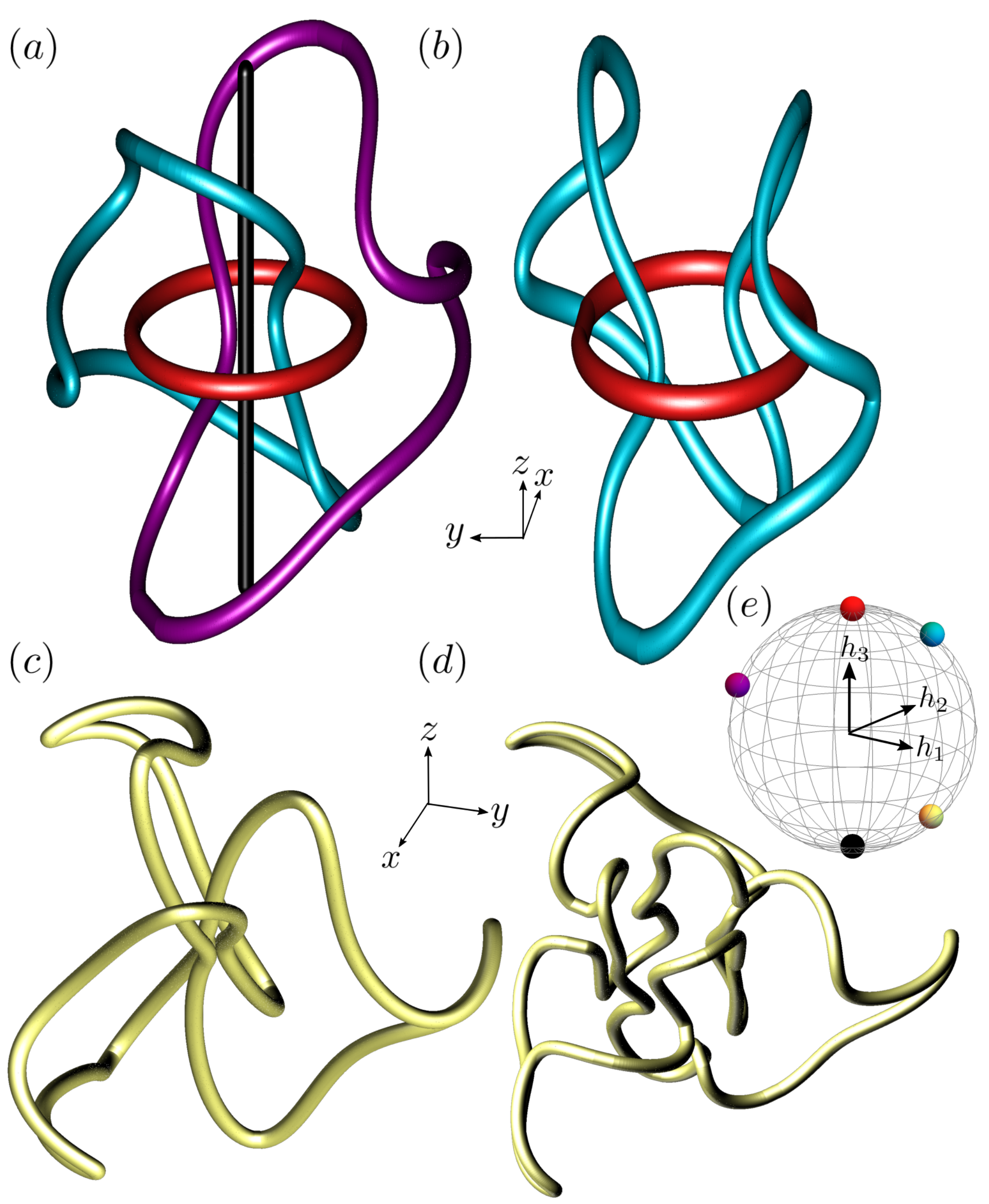}
	\vspace{-0.2cm}
	\caption{Links and knots in particle-like Hopfion optical excitations. (a-d) Real space preimages of Hopfions with different Hopf charge, $Q_H$, where each preimage corresponds to a point on the Poincar\'e sphere $S^2$ (e). Hopfion with (a) $Q_H=1$ (b) $Q_H=2$, where preimages interlink once and twice, respectively, and organize around the preimages of the vortex ring (red circle) and vortex line (black line corresponding to a circle of infinite radius). (c,d) $Q_H=6$ Hopfions with trefoil knot preimages. The Hopfions are created using the light field in Eq.~\eqref{Eq:SkyrmionField} with $l$ equal to (a) 1, (b) 2, (c,d) 3, and the $y$-polarized light component replaced in (c) by a two-photon transition and in (d) by the LG beams of Eq.~\eqref{Eq:Knot}. The beam widths $(w_x,w_1,w_2)/\lambda = (2,3,4.5)$.}
	\label{Fig:Hopfions}
\end{figure}

The topological structure of the Hopfion is revealed when considering the reduction in dimensionality of the parameter space under the Hopf map of Eq.~\eqref{Eq:HopfionMapping}, where multiple points on $S^3$ map to the same point on $S^2$.
These points form closed curves in real space, known as Hopfion preimages, which interlink an integer number of times, as the preimages of the Hopfion introduced in Fig.~\ref{Fig:ChargeDensity}(c) show in Fig.~\ref{Fig:Hopfions}(a).
The linking number is given by the Hopf charge, $Q_H$, which can be shown~\cite{Gudnason2020} to be equal to the 3D Skyrmion charge, Eq.~\eqref{Eq:3DSkyrmionNumber}.
Therefore, we can increase the preimage interlinking by increasing the total winding of vortex rings and lines, as discussed in Sec.~\ref{Skyrmions}.
Multiply-quantized vortex lines in $P_x$ can easily be prepared by changing the beam orbital angular momentum in Eq.~\eqref{Eq:SkyrmionField}. 
Choosing $l=2$, we form a $Q_H=2$ Hopfion with real space preimages that interlink twice, as shown in Fig.~\ref{Fig:Hopfions}(b).

Here we show how we can even prepare Hopfions that have the highly sought-after knotted structure, provided that the total winding of vortex rings is increased. This is more complicated than preparing higher quantized vortex lines, not least because multiply-quantized optical vortex rings are forbidden in paraxial light beams~\cite{berry_2001}. 
However, we overcome this limitation by two alternative strategies.
The first is to create the Hopfion using the field in Eq.~\eqref{Eq:SkyrmionField}, but where the $y$-polarized light component is now chosen to drive a two-photon transition, with the beam wavelength doubled. 
Each photon excites a single vortex ring, such that $P_y \propto [\text{U}_{0,0}(w_1)-c\text{U}_{0,0}(w_2)]^2$, therefore forming a doubly-quantized ring~\cite{Ruostekoski2005}, with $P_y \sim[(\rho-\rho_0)+\text{i}\zeta z]^2$ in the ring vicinity.
Using then $l=3$ for the LG beam in Eq.~\eqref{Eq:SkyrmionField} to prepare a triply-quantized vortex line that threads the vortex ring, we create a Hopfion that has a real space preimage of a trefoil knot, Fig.~\ref{Fig:Hopfions}(c).
An alternative method is to choose the $y$-polarized light component in Eq.~\eqref{Eq:SkyrmionField} to be structured according to the techniques of Refs.~\cite{Leach2004,berry_2001}, where using the superposition of LG modes,
\begin{equation}\label{Eq:Knot}
{\cal E}{}_y(\rv)\simeq 0.25\text{U}_{0,0}(w_x)-0.6\text{U}_{0,1}(w_x)+0.375\text{U}_{0,2}(w_x),
\end{equation}
gives a configuration of four coaxial vortex rings (two oppositely winding in the focal plane, one above and one below the focal plane), through which the triply-quantized vortex line threads and creates a Hopfion with a trefoil knot preimage, Fig.~\ref{Fig:Hopfions}(d).

\section{Singular defects}

Until now, we have considered non-singular topological textures where the orientation of the spin is well defined everywhere in space. We now show how it is also possible to form singular defects for which the spinor becomes ill-defined at a finite number of points. 
Structured light fields that exhibit singularities can create singular defects in atomic optical excitations by analogous principles to non-singular textures. 
To form 2D optical point defects in oblate atomic ensembles strongly confined along the $z$-direction, we consider a full Poincar\'e beam profile formed by a superposition of two LG beams with opposite orbital angular momenta~\cite{Donati2016},
\begin{equation}\label{vortex}
\boldsymbol{\mathcal{E}}{}(\textbf{r})=e^{-\text{i}\varphi/2}\text{U}_{1,0}(w_0)\hat{\textbf{e}}_++e^{\text{i}\varphi/2}\text{U}_{-1,0}(w_0)\hat{\textbf{e}}_-, 
\end{equation}
where $\varphi = 0$ ($\varphi = \pi$) results in an azimuthal (radial) singular vortex in the light field. 

\begin{figure}
	\hspace*{0cm}
	\includegraphics[width=\columnwidth]{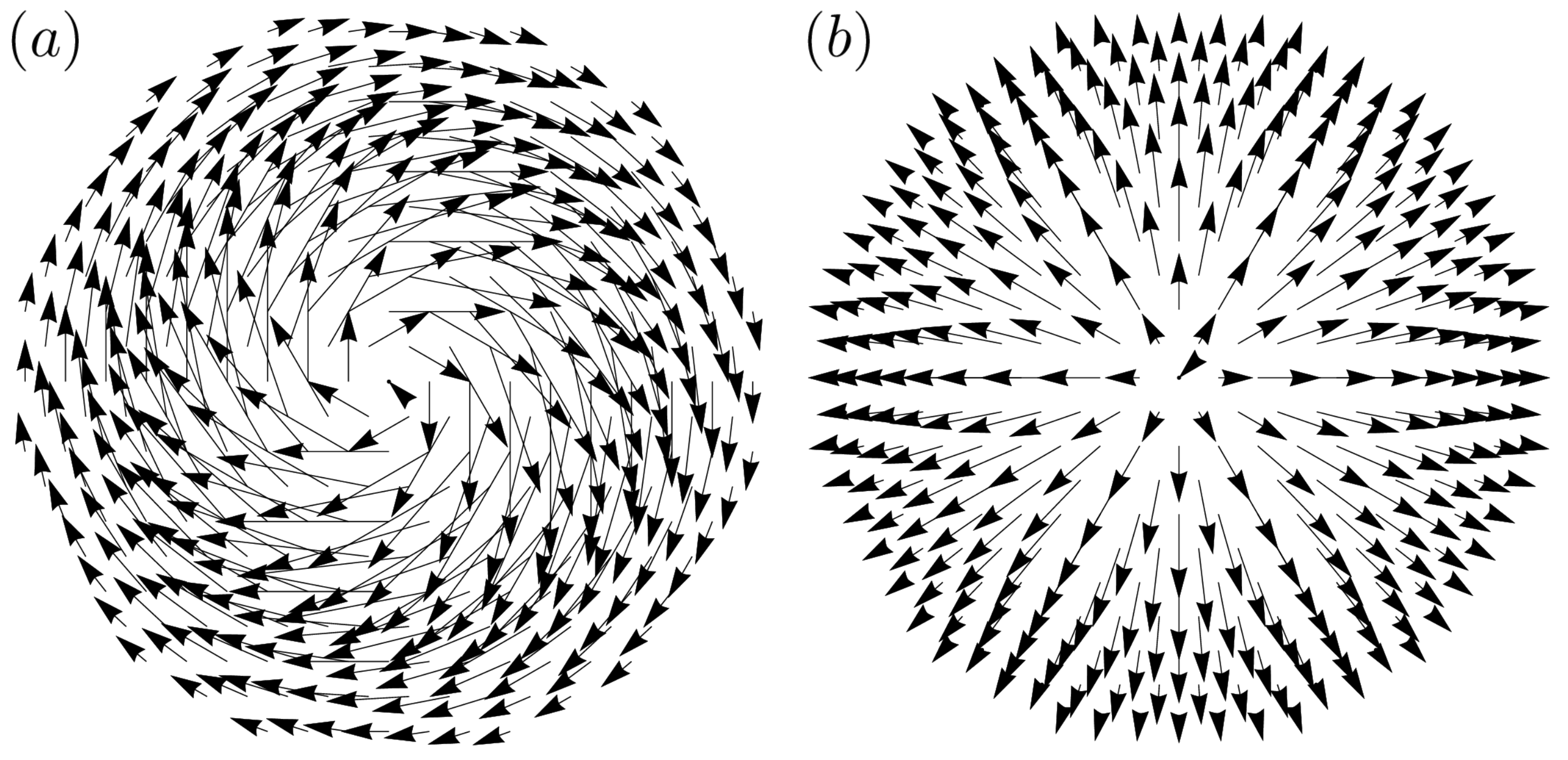}
	\vspace{-0.2cm}
	\caption{Collective optical excitations with singular-like defects. 
	(a) Azimuthal and (b) radial point vortex in the atomic polarization, generated in a steady-state response to the incident light field of Eq.~\eqref{vortex} that mediates dipole-dipole interactions between the atoms in an $N=313$ triangular array with circular boundaries and spacing $a/\lambda=0.5$. The beam width $w_0/\lambda = 2.77$, phase $\theta = 0$, and laser frequency detuning from the atomic resonance $\Delta/\gamma = 0.90$ for (a), or $\theta = \pi$ and $\Delta/\gamma = 0.89$ for (b). 
	}
	\label{Fig:Vortices2}
\end{figure}

For a dominant incident field, the singular configuration can be transferred onto the atomic polarization density without the need for any applied level shifts, as in Sec.~\ref{baby}. Such a configuration eventually radiates at the single-atom decay rate. 
However, remarkably, we find that specific defect structures can be highly robust and stable even in the strongly interacting limit where the incident field is no longer dominant in the atomic ensemble. These structures therefore represent spatially delocalized coherent `superatoms' that extend over the sample.  
In a cold and dense atomic ensemble, resonant incident light can scatter strongly, mediating  dipole-dipole interactions between the atoms. 
In Fig.~\ref{Fig:Vortices2}, we show the real components of the steady-state polarization density for interacting atoms driven by the field in Eq.~\eqref{vortex}, with $w_0/\lambda = 2.77$, in the limit of low light intensity where individual atoms respond to light as classical linear oscillators~\cite{Ruostekoski1997a, Lee16}.
For an atom spacing $a/\lambda=0.5$, the intensity of light scattered between nearest-neighbor atoms at the center of the lattice, $I_{\text{scat}}$, is much larger than the maximum intensity of the incident field, $I_{\text{inc}}$, with $I_{\text{scat}}/I_{\text{inc}} \simeq 2$. 
Therefore the atoms no longer emit light independently, but instead exhibit collective optical excitations, together with collective resonance linewidths and line shifts.
Despite the presence of strong collective behavior, the optical excitations in Fig.~\ref{Fig:Vortices2} show clear vortex-like structures similar to the incident field. To understand this behavior, we calculate the collective excitation eigenmodes of the interacting system.

We find that the system supports several collective eigenmodes with singular defects in the real components of the atomic polarization amplitudes. 
The resulting stationary excitations in Fig.~\ref{Fig:Vortices2} consist almost solely of a single collective excitation with an azimuthal (radial) defect, with a well-defined resonance linewidth and line shift, where the eigenmode occupation~\cite{Facchinetti18} reaches 99\% at the eigenmode resonance, $\Delta/\gamma = 0.90$ ($\Delta/\gamma = 0.89$).
$S^1\rightarrow S^1$ mappings determine the winding number (Poincar\'e index) for a singular topological defect, as the count of the net total change in the real components of the polarization density orientation around a closed loop, 
\begin{equation}\label{Eq:SingularWinding}
Q = \oint \frac{d\textbf{r}}{2\pi}\cdot \nabla \arctan\left(\frac{\hat{P}_y}{\hat{P}_x}\right),
\end{equation}
with $Q=1$ for the azimuthal and radial vortices.
In an infinite system, the collective eigenmodes are real, and the system exhibits true topological defects. 
However, for the small atomic ensembles considered here, the imaginary components of the eigenmodes do not entirely vanish, e.g., the azimuthal vortex eigenmode appearing in the stationary excitation of Fig.~\ref{Fig:Vortices2}(a) has a small 3\% contribution to the total polarization density amplitude from the imaginary part.
In comparison, the full stationary excitation has a 2\% contribution from the imaginary part.

\section{Concluding remarks}
3D particle-like topological objects have inspired research across a wide range of different disciplines. The ideas originate from Kelvin, who proposed how vortex strings forming closed loops, links, and knots could explain the structure of atoms~\cite{Thomson1869}. To transfer such universal concepts to light and optical excitations, standard textbook representations of field amplitudes in terms of the Stokes vector on the Poincar\'e sphere fall dramatically short of the goal. This is because particle-like topologies can only be achieved in the complete optical hypersphere description, where variation of the total electromagnetic phase of vibration is retained. Here we have constructed a comprehensive platform of topologically non-trivial optical excitations of atoms, induced by light. The resulting amplitudes of electronic vibrations have been shown to exhibit substantially more complex topologies than the incident light creating them. In addition, this allows topological objects to be stored in excitations in highly controllable quantum systems with long lifetimes.

The proposed setup potentially paves the way for applications in future quantum simulators.
The Skyrme model of 3D particle-like objects~\cite{Skyrme1961} is not only an elegant mathematical construction, but also simulates a low-energy limit of QCD where baryons are described by the quantised states of classical soliton solutions~\cite{ADKINS}. By describing these field configurations using linked Hopf maps, the particle-like objects take the form of links and knots, analogous to knotted solitons of the Skyrme-Faddeev model~\cite{faddeev_nature_1997, Battye98, Sutcliffe17, HIETARINTA99, Babaev2002}, and representing physical realisations of Kelvin’s ideas in optical excitations.

\section{Acknowledgements}
C.D.P.\ and J.R.\ acknowledge financial support from the UK EPSRC (Grant Nos.\ EP/S002952/1, EP/P026133/1), and M.R.D.\  from the EPSRC Centre for Doctoral Training in Topological Design (EP/S02297X/1).

%\bibliography{atomlightFirstNameTitle}

%apsrev4-2.bst 2019-01-14 (MD) hand-edited version of apsrev4-1.bst
%Control: key (0)
%Control: author (8) initials jnrlst
%Control: editor formatted (1) identically to author
%Control: production of article title (0) allowed
%Control: page (0) single
%Control: year (1) truncated
%Control: production of eprint (0) enabled
%

\end{document}